\documentclass[pra,twocolumn,showpacs]{revtex4}
\usepackage{epsfig}
\usepackage{amsbsy}
\usepackage{amsmath}
\usepackage{latexsym}

\begin{document}
\title{Entanglement oscillations in open Heisenberg chains}
\author{Ting Wang, Xiaoguang Wang, and Zhe Sun}
\affiliation{Zhejiang Institute of Modern Physics, Department of
Physics, Zhejiang University, HangZhou 310027, China}

\date{\today}  % It is always \today, today, but any date may be explicitly specified

\begin{abstract}
We study pairwise entanglements in spin-half and spin-one Heisenberg chains with an open boundary condition,
respectively. We find out that the ground-state and the first-excited-state entanglements are equal for the
three-site spin-one chain. When the number of sites $L>3$, the concurrences and negativities display oscillatory
behaviors, and the oscillations of the ground-state and the first-excited-state entanglements are out of phase
or in phase.
\end{abstract}
   % PACS, the Physics and Astronomy Classification Scheme.
\maketitle

\section{Introduction}
Recently, the study of entanglement properties in Heisenberg systems has received much
attention~\cite{M_Nielsen}-~\cite{Wang04}. Quantum entanglement is one of the most intriguing properties of
quantum physics~\cite{Einstein,schr} and the key ingredient of the emerging field of quantum information theory
and processing~\cite{Schliemann}, and can be exploited to accomplish some physical tasks such as quantum
teleportation~\cite{Tele}.

In most of the previous studies on entanglement of many-body
states, a periodic boundary condition is assumed for spin chains.
On the other hand, spin chains with an open boundary condition
(OBC) have been used to construct spin cluster
qubits~\cite{Meier1,Meier2} for quantum computation and employed
for quantum communication from one end to
another~\cite{Bose2,chris}. Perfect state transfer has been
obtained via the open chain without requiring qubit coupling to be
switched on and off~\cite{chris}. These investigations reveal that
open chains are of great advantage in implementing quantum
information tasks. Thus the study of the entanglement structure in
open spin chains will be of importance as the entanglement
underlies operations of quantum computing and quantum information
processing.

Most of the systems considered in precious studies are spin-half systems as there exists a good measure of
entanglement of two spin-halves, the concurrence~\cite{wootters}, which is applicable to an arbitrary state of
two spin halves. On the other hand, the entanglements in higher spin systems are not well-studied due to the
lack of good operational entanglement measures. Here, we will use the negativity~\cite{PH,Vidal} to investigate
entanglement in spin-one systems. In this paper, by using the concept of concurrence or negativity, we study
pairwise entanglement in spin-half and spin-one Heisenberg chains with an OBC, respectively.

The paper is organized as follows. In Sec.II, firstly, we give
exact results of the ground-state and first-excited-state pairwise
entanglements for the three-qubit and four-qubit spin-half open
Heisenberg model; and then, we present numerical results of the
corresponding entanglements for the $5\sim10$-qubit open
Heisenberg model; finally, we investigate open boundary effects on
thermal-state pairwise entanglement for $2\sim6$-qubit Heisenberg
model. In Sec.III, we study pairwise entanglement in spin-one open
Heisenberg chains. We conclude in Sec.IV.

\section{Spin-half system}
The Heisenberg Hamiltonian for the chain of $L$ qubits with an OBC
is given by
\begin{eqnarray}
   H & = & \sum_{i=1}^{L-1} J(2\mathbf{S}_i \cdot\mathbf{S}_{i+1}+\frac{1}{2})\nonumber \\
     & = & \sum_{i=1}^{L-1} J \mathcal{S}_{i,i+1},
\end{eqnarray}
where $\mathbf{S}_i$ is the spin-half operator for qubit $i$,
$\mathcal{S}_{i,i+1}=\frac{1}{2}
(1+\vec{\sigma}_{i}\cdot\vec{\sigma}_{i+1})$ is the swap operator
between qubit $i$ and $i+1$, and
$\vec{\sigma}_{i}=(\sigma_{ix},\sigma_{iy},\sigma_{iy})$ is the
vector of Pauli matrices. In the following discussion, we assume
$J=1$(antiferromagnetic case).

Due to the SU(2) symmetry in our Hamiltonian, the concurrence
quantifying the entanglement of two qubits is given
by~\cite{wangx,zsun}
\begin{equation}\label{eq:conc}
  C_{ij}=\max\{0,-2\langle\mathbf{S}_i\cdot\mathbf{S}_{j}\rangle-1/2\}=\max\{0,-\langle\mathcal{S}_{ij}\rangle\},
\end{equation}
we see that the entanglement is determined by the expectation
value of the swap operator.

In the three-qubit case, the Hamiltonian can be written as
\begin{eqnarray}
  H & = & \mathcal{S}_{12}+ \mathcal{S}_{23}\nonumber \\
    & = & 2(\mathbf{S}_{1}\cdot\mathbf{S}_{2}+\mathbf{S}_{2}\cdot\mathbf{S}_{3})+1\nonumber \\
    & = & (\mathbf{S}_{1}+\mathbf{S}_{2}+\mathbf{S}_{3})^2-\mathbf{S}_{1}^{2}-
    \mathbf{S}_{2}^{2}-\mathbf{S}_{3}^{2}-2\mathbf{S}_{1} \cdot\mathbf{S}_{3}+1\nonumber \\
    & = &
    (\mathbf{S}_{1}+\mathbf{S}_{2}+\mathbf{S}_{3})^2-(\mathbf{S}_{1}+\mathbf{S}_{3})^2-\mathbf{S}_{2}^{2}+1,
\end{eqnarray}
because $(\mathbf{S}_{1}+\mathbf{S}_{2}+\mathbf{S}_{3})^2$, $(\mathbf{S}_{1}+\mathbf{S}_{3})^2$ and
$\mathbf{S}_{2}^{2}$ commute with each other, we can use the standard angular momentum coupling theory to
calculate all the eigenvalues of this system: firstly, $\mathbf{S}_{1}$ couples with $\mathbf{S}_{3}$, then they
couple with $\mathbf{S}_{2}$ again. The results are
\begin{equation}\label{eq:e}
  E_{0}=-1(2), \quad E_{1}=1(2), \quad E_{3}=2(4),
\end{equation}
where the number in the bracket denotes the degeneracy. The ground state is analytically given by
\begin{eqnarray}
   |\Psi_{0}^{(1)}\rangle & = & \frac{1}{\sqrt{6}}(|001\rangle-2|010\rangle+|100\rangle),\nonumber \\
   \text{or} \quad |\Psi_{0}^{(2)}\rangle & = &\frac{1}{\sqrt{6}}(|110\rangle-2|101\rangle+|011\rangle),
\end{eqnarray}
and the first-excited state is
\begin{eqnarray}
   |\Psi_{1}^{(1)}\rangle & = & \frac{1}{\sqrt{2}}(|001\rangle-|100\rangle),\nonumber \\
   \text{or} \quad |\Psi_{1}^{(2)}\rangle & = &\frac{1}{\sqrt{2}}(|110\rangle-|011\rangle).
\end{eqnarray}

Because the three-qubit Hamiltonian has an exchange symmetry, namely, the Hamiltonian is invariant after
swapping qubits 1 and 3,
 $\langle\mathcal{S}_{12}\rangle=\langle\mathcal{S}_{23}\rangle=\frac{1}{2}\,\langle H\rangle=\frac{E}{2}$. Thus,
 from Eqs.(\ref{eq:conc}) and (\ref{eq:e}), the concurrence of two qubits in the
ground state is found to be
\begin{equation}
  C_{12}^{0}=1/2,
\end{equation}
and for the first-excited state, the concurrence is
\begin{equation}
  C_{12}^{1}=0.
\end{equation}
We see that the ground state is an entangled state, whereas the
first-excited state is not.

Now we consider the four-qubit case, the Hamiltonian is
\begin{equation}
  H=\mathcal{S}_{12}+\mathcal{S}_{23}+\mathcal{S}_{34},
\end{equation}
from which, we obtain all the eigenvalues of this system as
follows
\begin{align}
   E_{0}&=-\sqrt{3}(1), \, & E_{1}&=1-\sqrt{2}(3), \, &E_{3}&=1(3),\nonumber \\
E_{4}&=\sqrt{3}(1), \, & E_{5}&=1+\sqrt{2}(3), \, &E_{6}&=3(5),
\end{align}
Then, the ground state is
\begin{eqnarray}\label{gs}
|\Psi_{0}\rangle&=&\frac{1}{\sqrt{24+12\sqrt{3}}}[|0011\rangle-(2+\sqrt{3})|0101\rangle\nonumber\\
& &+(1+\sqrt{3})|0110\rangle +(1+\sqrt{3})|1001\rangle\nonumber\\
& & -(2+\sqrt{3})|1010\rangle+|1100\rangle],
\end{eqnarray}
and the first-excited state is
\begin{eqnarray}\label{first}
  |\Psi_{1}^{(1)}\rangle & = & \frac{1}{\sqrt{8+4\sqrt{2}}}[-|0001\rangle+(1+\sqrt{2})|0010\rangle\nonumber\\
   & &-(1+\sqrt{2})|0100\rangle+|1000\rangle],\nonumber \\
  \text{or}\quad |\Psi_{1}^{(2)}\rangle & = &\frac{1}{\sqrt{8+4\sqrt{2}}}[|0011\rangle-(1+\sqrt{2})|0101\rangle\nonumber\\
   & &+(1+\sqrt{2})|1010\rangle-|1100\rangle],\nonumber \\
   \text{or} \quad |\Psi_{1}^{(3)}\rangle & = & \frac{1}{\sqrt{8+4\sqrt{2}}}[-|0111\rangle+(1+\sqrt{2})|1011\rangle\nonumber\\
   & &-(1+\sqrt{2})|1101\rangle+|1110\rangle],
\end{eqnarray}

Thus, from Eqs.(\ref{eq:conc}), (\ref{gs})and (\ref{first}), we
get the concurrences of the ground state as
\begin{equation}
  C_{12}^{0}=C_{34}^{0}=\frac{3+2\sqrt{3}}{4+2\sqrt{3}}=0.8660 , \quad C_{23}^{0}=0,
\end{equation}
and for the first-excited state, the concurrences are
\begin{equation}
  C_{12}^{1}=C_{34}^{1}=0,\quad C_{23}^{1}=\frac{1+\sqrt{2}}{2+\sqrt{2}}=0.7071,
\end{equation}
From the analytical results of the concurrences of the ground state and the first excited state, we observe that
the concurrence oscillations emerge.

Then we calculate the nearest-neighbor concurrences for the cases
$L=5\sim10$ numerically, and the results are shown in Fig.~1.
\begin{figure}
\includegraphics[width=8cm]{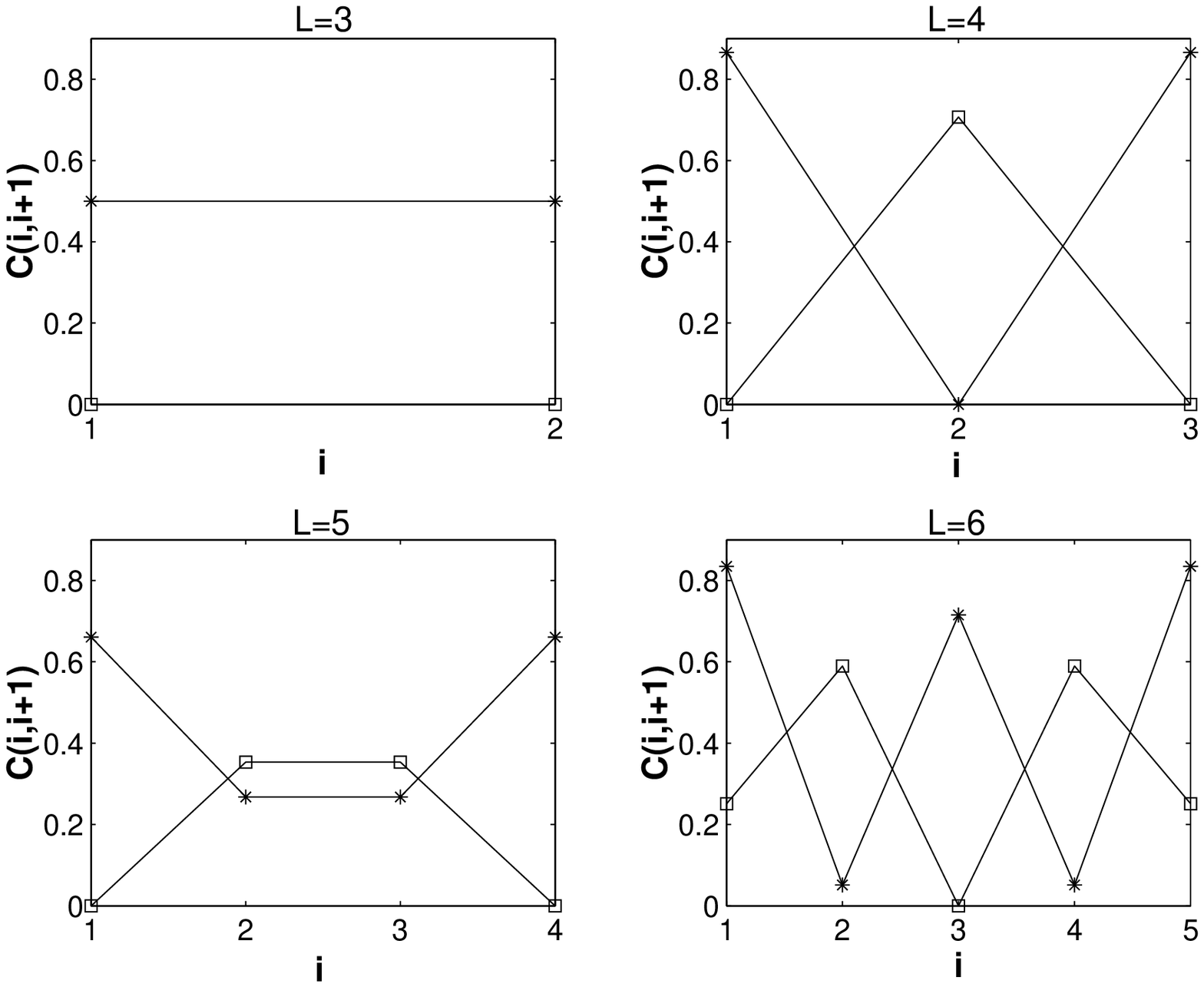}
\includegraphics[width=8cm]{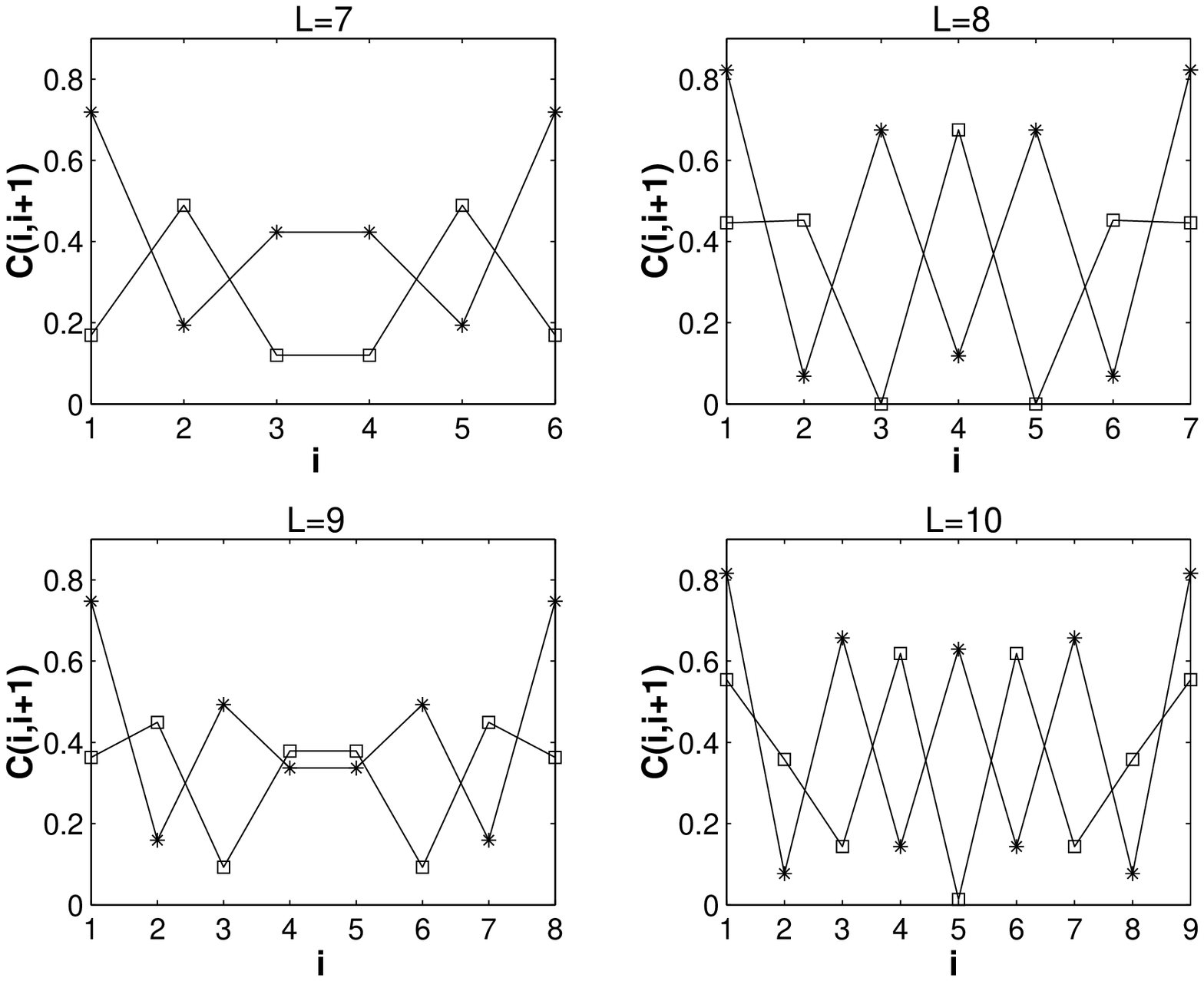}
\caption{\label{fig.1}The ground-state (star line) and
first-excited-state (square line) nearest-neighbor concurrences
versus site number for $L=3\sim10$ in the spin-half open
Heisenberg chains.}
\end{figure}
From Fig.1, we see that the ground-state and the first-excited-state concurrences oscillate when the site index
increases, and they are out of phase with each other when $L>3$ (in the case of $L=10$ , the upward convex of
the square line in the region of $i=1\sim3$ and $7\sim9$ is not so obvious). The reason of oscillatory behaviors
is as follows: for spin 2, there is a competition between spins 1 and 3, and they both favor being maximally
entangled with spin 2. If spin 2 shares a large entanglement with spin 1, then it is less entangled with spin 3,
and vice versa. Thus the oscillatory feature appears.

Now, we study entanglement in the thermal state, namely, consider the finite-temperature case. The state of a
system at thermal equilibrium is described by the density operator $\rho(T)=\exp(-\beta H)/Z$, where
$\beta=1/k_{B}T$, $k_{B}$ is the Boltzmann's constant, which is assumed to be 1 throughout the paper, and
$Z=\textrm{Tr}\{\exp(-\beta H)\}$ is the partition function. The entanglement in the thermal state is referred
as thermal entanglement. Due to $\langle\mathcal{S}_{ij}\rangle=\textrm{Tr}[\mathcal{S}_{ij}\cdot \rho]$, then
from Eq.(\ref{eq:conc}), the thermal concurrence quantifying the thermal entanglement is given by
\begin{equation}\label{eq:concT}
  C_{ij}(T)=\max\{0,-\textrm{Tr}[\mathcal{S}_{ij}\cdot \rho(T)]\},
\end{equation}
by using the above equation, the numerical results of the thermal concurrences  for $L=2\sim6$ are shown in
Fig.~2.
\begin{figure}
\includegraphics[width=8cm]{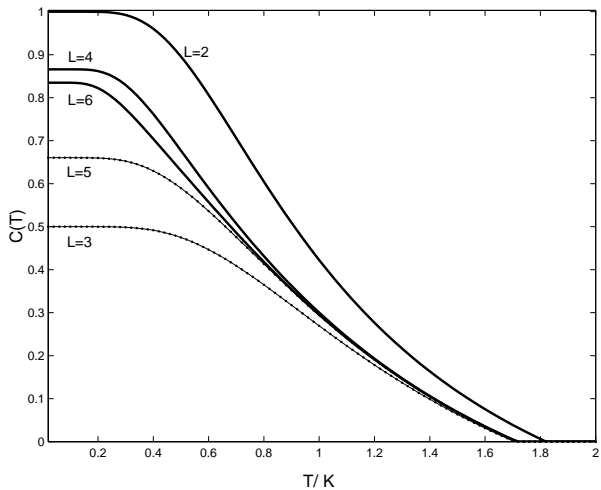}
\caption{\label{fig.2}The thermal concurrences between qubits 1 and 2 in the $2\sim6$-qubit spin-half open
Heisenberg chains.}
\end{figure}
We can see that the thermal entanglements for even qubits are larger than those for odd qubits. At a fixed lower
temperature, for even $L$ the thermal entanglements decrease as the number of qubits increases, whereas for odd
$L$, on the contrary, the entanglements increase as $L$ increases. As $L$ is very large, the effects of parity
of $L$ vanish. The threshold temperatures $T_{\text{th}}$s are independent of the qubits' number when $L$ is
large, and when the qubits' number is toward to infinity, they will approach a point, which is estimated as
$T_\text{th}=1.71658\sim1.71659K$.

\section{Spin-one system}
The Heisenberg Hamiltonian for the chain of $L$ spins in spin-one
system of an OBC is
\begin{equation}
  H  =  \sum_{i=1}^{L-1} J\mathbf{S}_i \cdot\mathbf{S}_{i+1},
\end{equation}
where $\mathbf{S}_i$ is the spin-one operator for spin $i$, $J$ is
the exchange constant. Here we also set $J=1$.

For the case of higher spins, a non-entangled state has necessarily a positive partial transpose (PPT) according
to the Peres-Horodecki criterion~\cite{PH}. In the case of two spin halves, and the case of (1/2,1) mixed spins,
a PPT is also sufficient. However, in the two spin-one particles, a PPT is not sufficient in general. The
two-spin state here displays a SU(2) symmetry, and for such a state, the PPT condition is necessary and
sufficient~\cite{breuer,Schliemann2}. The quantitative version of the criterion was developed by Vidal and
Werner~\cite{Vidal}. They presented a measure of entanglement called negativity that can be computed
efficiently, and the negativity does not increase under local manipulations of the system. So, we may use
negativity to exactly characterize the two-spin entanglement properties of our system.

For $L=3$ case, we can write the Hamiltonian that
\begin{eqnarray}
  H & = & \mathbf{S}_{1} \cdot\mathbf{S}_{2}+\mathbf{S}_{2}\cdot\mathbf{S}_{3}\nonumber \\
    & = &
    \frac{1}{2}[(\mathbf{S}_{1}+\mathbf{S}_{2}+\mathbf{S}_{3})^2-(\mathbf{S}_{1}+\mathbf{S}_{3})^2-\mathbf{S}_{2}^{2}],
\end{eqnarray}
using the similar method in the above section, all the eigenvalues of this system are directly obtained as
\begin{align}
   E_{0}&=-3(3), \quad & E_{1}&=-2(1), \quad &E_{3}&=-1(8),\nonumber \\
E_{4}&=0(3), \quad & E_{5}&=1(5), \quad &E_{6}&=2(7).
\end{align}
The ground state is obtained as
\begin{eqnarray}
  |\Psi_{0}^{(1)}\rangle & = & \frac{1}{\sqrt{60}}(|002\rangle+6|020\rangle+|200\rangle-3|011\rangle\nonumber\\
   & &+2|101\rangle-3|110\rangle),\nonumber \\
  \text{or}\quad |\Psi_{0}^{(2)}\rangle & = &\frac{1}{\sqrt{60}}(-2|012\rangle+3|102\rangle+3|120\rangle+3|021\rangle\nonumber\\
   & &+3|201\rangle-2|210\rangle-4|111\rangle),\nonumber \\
   \text{or} \quad |\Psi_{0}^{(3)}\rangle & = & \frac{1}{\sqrt{60}}(|022\rangle+6|202\rangle+|220\rangle-3|112\rangle\nonumber\\
   & &+2|121\rangle-3|211\rangle),
\end{eqnarray}
and the first-excited state is
\begin{eqnarray}
   |\Psi_{1}\rangle & = &\frac{1}{\sqrt{6}}(-|012\rangle+|102\rangle-|120\rangle\nonumber\\
   & &+|021\rangle-|201\rangle+|210\rangle).
\end{eqnarray}

In spin-one system, the negativity quantifying the pairwise
entanglement is~\cite{wangli}
\begin{eqnarray}
  \mathcal{N}_{ij}& = &\frac{1}{2}\,\max[0,\langle\mathcal{S}_{ij}\rangle-\langle\mathbf{S}_{i}\cdot\mathbf{S}_{j}\rangle-1]\nonumber\\
& &+\frac{1}{3}\,\max[0,-\langle\mathcal{S}_{ij}\rangle],
\end{eqnarray}
where $\mathcal{S}_{ij}=\mathbf{S}_{i}\cdot\mathbf{S}_{j}+(\mathbf{S}_i\cdot\mathbf{S}_{j})^2-I$ is the swap
operator between the spin-one particles $i$ and $j$. For the Hamiltonian of $L=3$, the spins 1 and 3 have an
exchange symmetry which leads to
$\langle\mathbf{S}_{1}\cdot\mathbf{S}_{2}\rangle=\langle\mathbf{S}_{2}\cdot\mathbf{S}_{3}\rangle$ and
$\langle\mathcal{S}_{12}\rangle=\langle\mathcal{S}_{23}\rangle$, so we have $\mathcal{N}_{12}=\mathcal{N}_{23}$.

Thus, from Eqs.(19) and (20), the expectation values of
$\mathcal{S}_{12}$ and $\mathbf{S}_{1}\cdot\mathbf{S}_{2}$ in the
ground state are given by
\begin{equation}
\langle\mathcal{S}_{12}\rangle^{0}=\frac{1}{6}, \quad
\langle\mathbf{S}_{i}\cdot\mathbf{S}_{j}\rangle^{0}=-\frac{3}{2},
\end{equation}
and in the first-excited state, they are
\begin{equation}
\langle\mathcal{S}_{12}\rangle^{1}=-1, \quad
\langle\mathbf{S}_{i}\cdot\mathbf{S}_{j}\rangle^{1}=-1.
\end{equation}
Then, from Eq.(21), we obtain the corresponding negativities as
\begin{equation}
  \mathcal{N}_{12}^{0}=1/3, \quad \mathcal{N}_{12}^{1}=1/3.
\end{equation}
From the above equation, we can see that the negativities in the ground state and the first-excited state are
equal. When $L>3$, the negativities in the ground state and the first-excited state are not equal, so this
equality is a mathematical accident. In addition, because the reduced density matrices in the ground state and
the first-excited state are different.
\begin{figure}
\includegraphics[width=8cm]{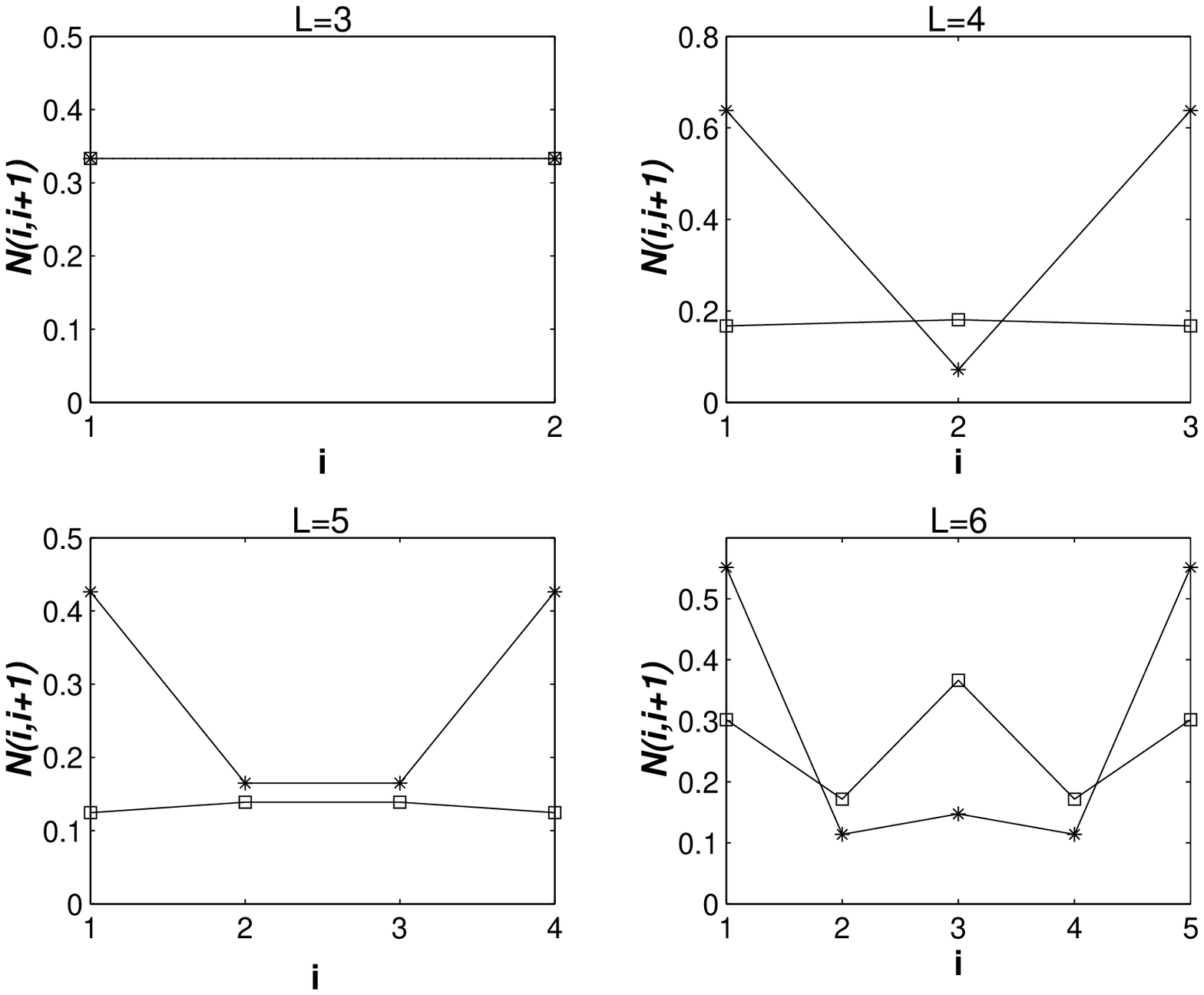}
\caption{\label{fig.3}The ground-state (star line) and first-excited-state (square line) nearest-neighbor
negativities versus site number for $L=3\sim6$ in the spin-one open Heisenberg chains.}
\end{figure}
\begin{figure}
\includegraphics[width=8cm]{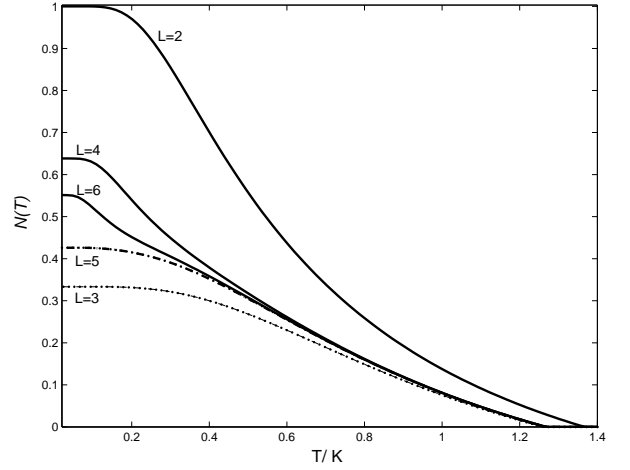}
\caption{\label{fig.4}The thermal negativities between spins 1 and 2 in the $2\sim6$-spin spin-one open
Heisenberg chains.}
\end{figure}

Numerically we calculate the negativities of $L=4\sim6$, the results are given in Fig.3. We can see that the
negativities oscillate when $L>3$, and the ground-state and the first-excited-state negativities are out of
phase with each other for $L=4$ and 5 cases, but they are in phase for $L=6$. Then from Fig.1 and Fig.3,
considering the problem of the oscillatory phase of the ground-state and the first-excited-state entanglements,
we find out that: on the edge of the Heisenberg chains, due to the effect of the boundary condition, the
ground-state entanglements are biggest, thus the oscillatory phases of the ground-state entanglements are
independent of the spins' number and invariant; but for the first-excited-state entanglements, the entanglements
on the edge are small when the spins' number is not big, then because of the existence of the competition, the
nearest entanglements will be large, thus the ground-state and the first-excited-state entanglements are out of
phase with each other; along with the spins' number increasing, the first-excited-state entanglements on the
edge will be larger in contrast with the corresponding ground-state entanglements, which leads to the nearest
first-excited-state entanglements diminishing, thus when the chain has enough spins, the ground-state and the
first-excited-state entanglements are in phase.

Becase
$\langle\mathbf{S}_i\cdot\mathbf{S}_{j}\rangle=\textrm{Tr}[(\mathbf{S}_i\cdot\mathbf{S}_{j})\cdot
\rho]$,$\langle(\mathbf{S}_i\cdot\mathbf{S}_{j})^{2}\rangle=\textrm{Tr}[(\mathbf{S}_i\cdot\mathbf{S}_{j})^{2}\cdot\rho]$,
then from Eq.(21), the thermal negativity quantifying the thermal
entanglement is obtained as
\begin{eqnarray}\label{thermal}
  \mathcal{N}_{ij}(T)& = &\frac{1}{2}\,\max\{0,\textrm{Tr}[(\mathbf{S}_i\cdot\mathbf{S}_{j})^{2}\cdot\rho(T)]-2\}\nonumber\\
  & &+\frac{1}{3}\,\max\{0,1-\textrm{Tr}[(\mathbf{S}_i\cdot\mathbf{S}_{j})\cdot\rho(T)]\nonumber\\
  & &-\textrm{Tr}[(\mathbf{S}_i\cdot\mathbf{S}_{j})^{2}\cdot\rho(T)]\},
\end{eqnarray}
according to Eq.(\ref{thermal}), we calculate the thermal negativities of $L=2\sim6$ numerically, which are
shown in Fig.4. Comparing Fig.4 with Fig.2, we  find out that the behaviors of thermal entanglement in spin-one
system is similar to those in spin-half system, but the threshold temperatures in spin-one system are all
smaller than those in spin-half system, and when the number of spins tends to infinity, they will approach a
point which is in the region of $T_{\text{th}}=1.26753\sim1.26758K$.

\section{conclusion}
In the above, we have studied pairwise entanglements in spin-half and spin-one Heisenberg chains with an OBC,
respectively. Some analytical and numerical results of the ground-state and first-excited-state pairwise
entanglements in the chains for several spin particles have been presented, and we find out some interesting
results: the ground-state and the first-excited-state negativities are equal when $L=3$; and when $L>3$, the
concurrences and negativities both oscillate, which results from the competition between the two spins on both
sides of each spin. We also have given numerical results of the thermal-state pairwise entanglements in the
chains for a few spins, the results reveal that the thermal entanglements in the two systems are similar and the
threshold temperatures in the two systems are both independent of the spins' number when the number is large,
and they will both approach a point when the number tends towards infinity. It is also interesting to study
entanglement of highly-exited eigenstates, which are under consideration.

\begin{acknowledgments}
This work is supported by NSF-China under grant No.10405019, Specialized Research Fund for the Doctoral Program
of Higher Education (SRFDP) under grant No.20050335087, and the project is also sponsored by SRF for ROCS and
SEM.
\end{acknowledgments}


\begin{thebibliography}{99}
\bibitem{M_Nielsen}  M. A. Nielsen, Ph. D thesis, University of Mexico,
1998, quant-ph/0011036;

\bibitem{Bose}  S. Bose and V. Vedral, \pra \textbf{61}, 040101 (2000).

\bibitem{M_Arnesen}  M. C. Arnesen, S. Bose, and V. Vedral, Phys. Rev. Lett.
\textbf{87}, 017901 (2001).

\bibitem{M_Gunlycke}  D. Gunlycke, V. M. Kendon, V. Vedral, and S. Bose, %
\pra \textbf{64}, 042302 (2001).

\bibitem{M_OConnor}  K. M. O'Connor and W. K. Wootters, \pra \textbf{63},
0520302 (2001).

\bibitem{M_Meyer}  D. A. Meyer and N. R. Wallach, quant-ph/0108104.

\bibitem{M_Wang01}  X. Wang, Phys. Rev. A \textbf{64}, 012313 (2001); Phys.
Lett. A \textbf{281}, 101 (2001); X. Wang and P. Zanardi, Phys.
Lett. A \textbf{301}, 1 (2002); X. Wang, Phys. Rev. A \textbf{66},
044305 (2002); X. Wang, H. Fu, and A. I. Solomon, J. Phys. A:
Math. Gen. \textbf{34}, 11307(2001); X. Wang and K. M\o lmer, Eur.
Phys. J. D \textbf{18}, 385(2002).

\bibitem{M_Kamta}  G. L. Kamta and A. F. Starace, Phys. Rev. Lett. \textbf{88}, 107901 (2002).

\bibitem{M_Osborne}  T. J. Osborne and M. A. Nielsen, \pra \textbf{66},
032110 (2002).

\bibitem{M_Osterloh}  A. Osterloh, L. Amico, G. Falci and R. Fazio, Nature
\textbf{416}, 608 (2002).

\bibitem{M_Yeo}  Y. Yeo, \pra \textbf{66}, 062312 (2002).

\bibitem{Jaeger}  G. Jaeger, A. V. Sergienko, B. E. A. Saleh, and M. C.
Teich, \pra \textbf{68}, 022318 (2003).

\bibitem{M_Sun}  Y. Sun, Y. G. Chen, and H. Chen, Phys. Rev. A \textbf{68},
044301 (2003).

\bibitem{Glaser}  U. Glaser, H. B\"{u}ttner, and H. Fehske, \pra \textbf{68}%
, 032318 (2003).

\bibitem{M_Santos}  L. F. Santos, \pra \textbf{67}, 062306 (2003).

\bibitem{M_Khv}  D. V. Khveshchenko, \prb \textbf{68}, 193307 (2003).

\bibitem{M_Zhou}  L. Zhou, H. S. Song, Y. Q. Guo, and C. Li, \pra \textbf{68}
, 024301 (2003).

\bibitem{Exp}  S. Ghose, T. F. Rosenbaum, G. Aeppli, and S. N. Coppersmith,
Nature (London) \textbf{425}, 48 (2003).

\bibitem{Gu}  S. J. Gu, H. Q. Lin, and Y. Q. Li, \pra \textbf{68}, 042330
(2003).

\bibitem{QPT_GVidal}  G. Vidal, J. I. Latorre, E. Rico, and A. Kitaev, Phys.
Rev. Lett. \textbf{90}, 227902 (2003).

\bibitem{M_Brennen}  G. K. Brennen, S. S. Bullock, \pra \textbf{70}, 52303
(2004).

\bibitem{Suncp}  R. Xin, Z. Song, and C. P. Sun, quant-ph/0411177.

\bibitem{Cirac1}  F. Verstraete, M. Popp, and J. I. Cirac, Phys. Rev. Lett.
\textbf{92}, 027901 (2004).

\bibitem{Cirac2}  F. Verstraete, M. A. Mart\'{i}n-Delgado, J. I. Cirac, Phys.
Rev. Lett. \textbf{92}, 087201 (2004).

\bibitem{QPT_JVidal}  J. Vidal, G. Palacios, and R. Mosseri, Phys. Rev. A
\textbf{69}, 022107 (2004).

\bibitem{QPT_Lambert}  N. Lambert, C. Emary, and T. Brandes, Phys. Rev.
Lett. \textbf{92}, 073602 (2004).

\bibitem{Fan}  H. Fan, V. Korepin, and V. Roychowdhury, \prl \textbf{93},
227203 (2004).

\bibitem{Wang04}X. Wang, Phys. Lett. A \textbf{329}, 439 (2004); Phys. Lett. A, \textbf{334}, 352 (2005).

\bibitem{Einstein}A. Einstein, B. Podolsky, and N. Rosen, Phys.
Rev. {\bf 47}, 777 (1935).

\bibitem{schr}E. Schr\"{o}dinger, Naturwissenschaften {\bf 23},
807 (1935).

\bibitem{Schliemann}  J. Schliemann, \pra \textbf{68}, 012309 (2003).

\bibitem{Tele}  C. H. Bennett, G. Brassard, C. Crepeau, R. Jozsa, A. Peres,
and W. K. Wootters, \prl \textbf{70}, 1895 (1993).

\bibitem{Meier1}F. Meier, J. Levy, and D. Loss, \prl \textbf{90},
047901 (2003).

\bibitem{Meier2}F. Meier, J. Levy, and D. Loss, \prb \textbf{68},
134417 (2003).

\bibitem{Bose2}S. Bose, \prl \textbf{91}, 207901(2003); V.
Subrahmanyam, \pra \textbf{69}, 034304 (2004).

\bibitem{chris}M. Christandl, N. Datta, A. Ekert, and A. J.
Landahl, \prl \textbf{92}, 187902 (2004).

\bibitem{wootters}S. Hill and W. K. Wootters, \prl \textbf{78},
5022 (1997); W. K. Wootters, \prl \textbf{80}, 2245 (1998).

\bibitem{PH}A. Peres, \prl {\bf 77}, 1413 (1996);
M. Horodecki, P. Horodecki, and R. Horodecki, Phys. Lett. A {\bf
223}, 1 (1996).

\bibitem{Vidal}  G. Vidal and R. F. Werner, \pra \textbf{65}, 032314 (2002).

\bibitem{wangx}X. Wang, Phys. Rev. E {\bf 69}, 066118 (2004).

\bibitem{zsun}Z. Sun, X. Wang, A. Z. Hu, and Y. Q. Li, Commun. Theor. Phys. (Beijing, China) {\bf 43} (2005) pp.
1033 - 1036.

\bibitem{breuer} H. P. Breuer, Phys. Rev. A \textbf{71}, 062330 (2005).

\bibitem{Schliemann2}J. Schliemann, Phys. Rev. A 72, 012307 (2005).

\bibitem{wangli}X. Wang, H. B. Li, Z. Sun, and Y. Q. Li, J. Phys. A: Math. Gen. {\bf 38} (2005) 8703.
\end{thebibliography}
\end{document}